\pdfoutput=1
\UseRawInputEncoding 
\documentclass[journal,onecolumn]{IEEEtran}
\usepackage{graphicx}   
\usepackage{epstopdf}
\usepackage{physics}
\usepackage{amsmath,amssymb}
\usepackage[english, footnote, final, nomargin, marginclue]{fixme}
\fxusetheme{color}
\usepackage[colorlinks=true, pdfstartview=FitV, linkcolor=blue, citecolor=blue, urlcolor=blue]{hyperref}
\usepackage{float}
\restylefloat{table}
\usepackage{diagbox}
\usepackage{multirow}
\usepackage{array}
\usepackage[export]{adjustbox}

\begin{document}
\bstctlcite{BSTcontrol}
\title{A Synergy of Institutional Incentives and Networked Structures in 
Evolutionary Game Dynamics of Multi-agent Systems
 }

\author{Ik~Soo~Lim,
Valerio~Capraro
\thanks{I.\,S.\,Lim is with 
School of Computing and Mathematical Sciences,
University of Greenwich, London, UK
(e-mail: i.lim@gre.ac.uk).
\\
V.\,Capraro is with
Economics Department, Middlesex University London, UK.
}
}

\markboth{
}%
{Shell \MakeLowercase{\textit{et al.}}: Bare Demo of IEEEtran.cls for IEEE Journals}

\maketitle

\begin{abstract}
Understanding the emergence 
of prosocial behaviours (e.g.,\,cooperation and trust) among self-interested agents is an important problem in many disciplines.
Network structure and institutional incentives (e.g., punishing antisocial agents) are known to promote prosocial behaviours, when acting in isolation, 
one mechanism being present at a time.
Here we study the interplay between these two mechanisms
to see whether they are independent, interfering or synergetic.
Using evolutionary game theory,
we show that 
punishing antisocial agents and a regular networked structure
not only promote prosocial behaviours among agents playing the trust game, 
but they also interplay with each other, leading to interference or synergy, depending on the game parameters. 
Synergy emerges on a wider range of parameters than interference does. 
In this domain, 
the combination of incentives and networked structure improves the efficiency of incentives, yielding prosocial behaviours at a lower cost than the incentive does alone.
This has a significant implication in the promotion of prosocial behaviours in multi-agent systems.
\end{abstract}

\begin{IEEEkeywords}
Game theory, multi-agent system, evolutionary dynamics, trust game, synergy.
\end{IEEEkeywords}

\IEEEpeerreviewmaketitle

\section{Introduction}

In a wide range of disciplines,
it is a fundamental challenge 
to understand the emergence and maintenance of prosocial behaviours among self-interested agents 
\cite{Zhang:2020aa}%
--\cite{Sigmund:2010aa}.
Evolutionary game theory is widely used to study game dynamics of multi-agent systems
in games involving prosociality, as
for instance,
the prisoner's dilemma,
the public goods game,
the labour division, 
and the trust game
\cite{nowak2006five,Zhang:2020aa,Zhang:2019aa,Abbass:2016aa}.
Specifically,
the replicator equations are widely used for evolutionary game dynamics of multi-agent systems
in well-mixed populations,
where more successful behaviours are preferably imitated via social learning
\cite{taylor1978evolutionary,Zhang:2020aa,Zhang:2019aa,Sasaki:2012rz}.
Without any additional mechanism,
evolutionary game dynamics drives agents not to behave prosocially
and, thus, they end up with a lower payoff
than what they would get if they all behave prosocially.
       
Various mechanisms have been proposed to promote prosocial behaviours
such as incentives
and network 
reciprocity
\cite{nowak2006five,Zhang:2020aa}.
Network reciprocity yields the evolution of prosocial behaviours by self-organised clusters of prosocial agents:
the spatial structure
constrains agents to interact with and imitate only (immediately) neighbouring agents
\cite{nowak2006five,Ohtsuki:2006fk}%
--\cite{Pacheco:2006ve}.
Institutional incentives lead to the evolution of prosocial behaviours
as well,
by either penalising agents for antisocial behaviours
or rewarding for prosocial behaviours \cite{Zhang:2020aa,Sasaki:2012rz,Sigmund:2010aa,Fang:2021ut}. 
Previous works have often studied these two mechanisms in isolation, missing potential interplays between them. In this paper we move a first step towards filling this gap. 

Specifically,
we use a variant of the replicator equations for evolutionary game dynamics of agents playing the 
trust game 
in a structured population with institutional incentives.
We analyse
the interplay between 
these two mechanisms in the domain of 
regular graphs
to see whether a combination of them 
favour or disfavours the evolution of
prosocial behaviours.
We also find the optimal level of incentive that maximises the population payoff or social welfare,
considering the operating cost of incentives. 
Most previous works have focused on which incentives promote prosocial behaviours best.
Another useful measurement for the success of an incentive would be the payoff 
at the evolutionarily stable state \cite{Dong:2019aa}.

We decided to focus on the trust game
(TG) that
has been widely used to formally 
study trust and trustworthiness in various disciplines
\cite{Abbass:2016aa,Lim:2020aa},
\cite{Johnson:2011aa}%
--\cite{Masuda:2012aa}. 
The TG is a prototypical game or an abstraction of economic transactions that involve buyer and seller interactions for a product or service.
Trusting others and reciprocating trust with trustworthy behaviours are important elements of
successful economic and social interactions \cite{Johnson:2011aa}.
In engineering research communities,
the concept of trust has also attracted significant interest, 
ranging from networking to human-machine interaction 
and
artificial intelligence \cite{Cho:2015aa}
while many problems are cast as buyer-seller 
scenarios \cite{Jung:2019yq}, 
\cite{Niyato:2009dq}.

\section{Model: The Symmetric Binary Trust Game}

We use a variant of 
the TG
 \cite{Masuda:2012aa}, 
which makes it a stronger social dilemma than the conventional 
TG 
\cite{Abbass:2016aa}, \cite{Johnson:2011aa}.
In our variant of the TG, there are two agents, an investor and a trustee.
The investor first decides whether to invest in the trustee or not.
If the investor does not invest, then both agents receive a payoff of 0.
If the investor invests, then the trustee decides whether to behave trustworthily or untrustworthily.
If the trustee is trustworthy (i.e.,\,s/he shares the gain stemming from the investment with the investor)
both agents receive 
$r$, where $0<r<1$.
If the trustee is untrustworthy 
(i.e.,\,not sharing it
with the investor),
the trustee gets $1$
while the investor gets $-1$.  
Given a trusting investor,
the TG is a social dilemma 
because: 
(i) the amount of total payoff ($2r$ or $0$) depends on the trustee's strategy
and 
(ii) the trustee maximises their payoff by being untrustworthy, 
 which yields a higher payoff of 1 to  the trustee
  but a lower total payoff of 0.

The TG in its original form is asymmetric, 
meaning that
an agent exclusively plays either as an investor or as a trustee
\cite{Masuda:2012aa}.
We consider a symmetric TG
such that given a pair of agents,
one plays as an investor and the other as a trustee:
the role allocation is randomly determined with equal probability of 1/2.
As an investor, one either invests in a trustee or not.
As a trustee, one either acts trustworthily  or untrustworthily.
Hence, there are 4 strategies that an agent can take $\{IT, IU, NT, NU\}$,
where $I$ and $N$ respectively denote `invest' and `not invest',
while $T$ and $U$ respectively denote `trustworthy' and `untrustworthy'.
The payoff matrix of the symmetric TG 
 is given
 (up to the factor 1/2, which we hereafter omit)
 by
\begin{equation}
\Pi=
 \bordermatrix{
       & IT & IU & NT & NU \cr
     IT&  2r &  -1 +r&   r & -1  \cr
    IU &  r +1& 0 & r    & -1  \cr
     NT &  r &   r &  0 & 0  \cr
     NU &  1 & 1   &   0& 0  
  },
\end{equation}
where  the elements denote 
the
payoffs that 
an agent
adopting the strategies in the rows acquires when interacting with 
an agent adopting the strategies in the columns.
For instance, the payoff 
of
an agent playing $IT$ with an agent  playing $IU$ is $-1 +r$.

\subsection{Incentives}
To promote prosocial behaviours (i.e.,\,$I$ and $T$),
an institutional incentive scheme 
lowers the payoff of an agent who acts untrustworthily as a trustee
towards
an investing investor.
The payoff matrix due to the penalty is given by
\begin{equation}
P =
          \bordermatrix{
               & IT & IU & NT & NU \cr
             IT&  0 &   0&   0& 0  \cr
            IU &  -p& -p& 0    & 0  \cr
             NT &  0 &   0 &  0 & 0  \cr
             NU &  -p & -p   &   0& 0  
          }, 
\end{equation}
where $p \ge 0$ is the expected fine.
We assume that each agent pays a tax $f \ge 0$ to maintain the incentive-providing institution.  
The payoff matrix due to the tax is given by $F = -f J_4$,
where $J_4$ is a 4$\times$4 matrix with every element being 1.
Hence, the net payoff matrix $A$ is given by
\begin{equation}
A = \Pi+ P +F.
\end{equation}
 
\subsection{Evolutionary Game Dynamics}

\begin{figure*}[t!]  
\begin{center}        
\includegraphics[width=0.99\textwidth]{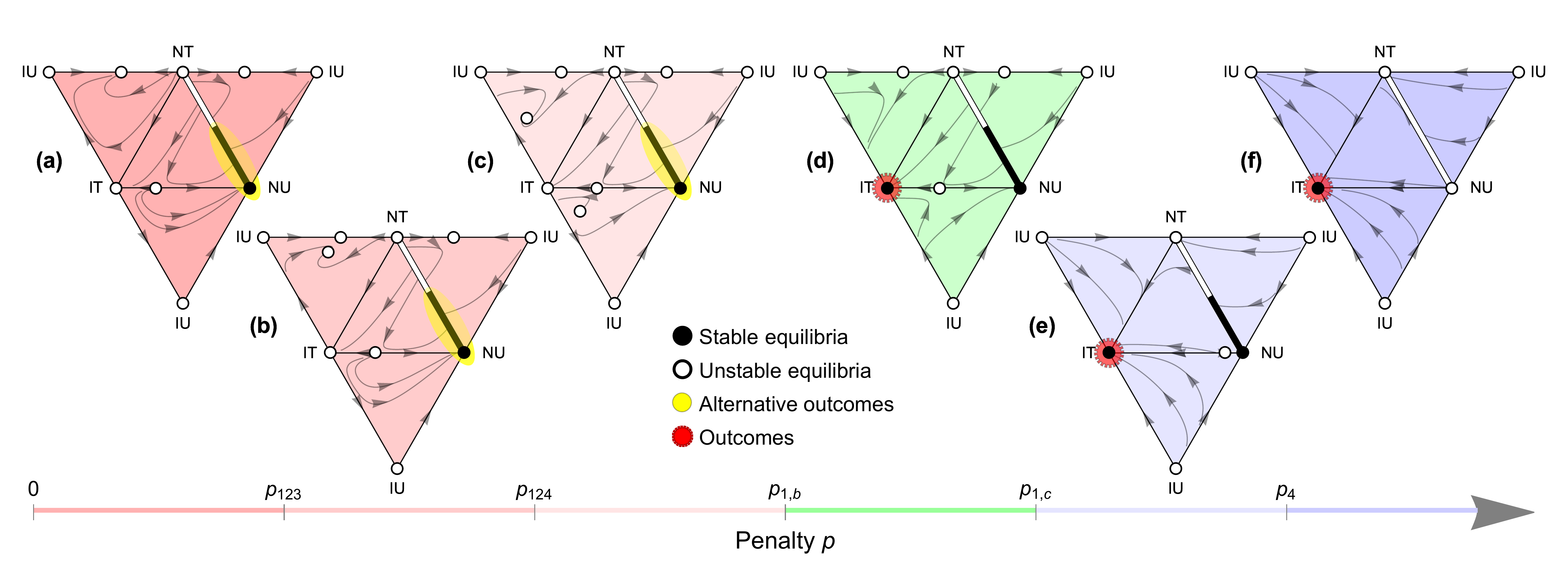}          
\end{center}  
\vspace{-0.3cm}
\caption{
The effect of punishment and 
networked structure (i.e.\,a random regular graph)
on the evolutionary dynamics of the TG  in terms of 
 penalty
 size $p$.
The triangles are the boundaries of a
3-simplex
representing the state space.
If the penalty is small $p <p_{1,b}$,
the population state spends 
most of the time
on the stable part of the line of equilibria $NT$--$NU$ \textbf{(a)} to \textbf{(c)}.
If $p \ge p_{1,b}$, 
the population state converges to the $IT$ vertex, i.e.,\,full trust and trustworthiness
\textbf{(d)} to \textbf{(f)}.
\textbf{(a)} For $p <p_{123}$, equilibria appear only on the vertices and edges.
\textbf{(b)} For $p_{123} <p <p_{124}$,
as $p$ crosses $p_{123}$,
the equilibrium in the $IT$--$IU$--$NT$ face emerges from the equilibrium on the $IU$--$NT$ edge.
Although the equilibrium in the  
face has two negative (real parts of) eigenvalues,
it is unstable 
since the remaining eigenvalue is positive.
\textbf{(c)} For $p_{124} <p <p_{1,b}$,
as $p$ crosses $p_{124}$,
 the equilibrium in the  $IT$--$IU$--$NU$ face emerges from the equilibrium on the $IT$--$NU$ edge.
\textbf{(d)} For $p_{1,b} <p <p_{1,c}$,
as $p$ crosses $p_{1,b}$,
the equilibria in the faces exit the simplex through the $IT$--$IU$ edge and
the $IT$ vertex is globally asymptotically stable:
from any initial conditions, trajectories converge to $IT$.
Even a population state initially on the stable part of the line of equilibria eventually
converges to $IT$.
The state fluctuates along the line by neutral drift due to random perturbations.
Once it has reached the unstable part of the line,
an arbitrary small random perturbation can drive it to $IT$.
\textbf{(e)} For $p_{1,c} <p <p_4$,
as $p$ crosses 
the value $p_{1,c}$, which
would be required for the evolution of $IT$ in a well-mixed population, 
the equilibrium on the $IU$--$NT$ edge moves toward the $NT$ vertex.
As $p$ increases and crosses $p_{23}$,
the equilibrium merges with the $NT$ vertex.
As $p$ further increases,
the equilibrium on the $IT$--$NU$ edge moves toward the $NU$ vertex
and the stable part of the line of equilibria shrinks.
\textbf{(f)} For $p_4 <p$,
as $p$ crosses $p_4$,
the equilibrium on the $IT$--$NU$ edge merges with the $NU$ vertex
and the whole line of equilibria becomes unstable as well as the $NU$ vertex.
Parameters: $r=0.6, k=10$,
$p/p_{1,c} =p/(1-r) = $ 0.2 \textbf{(a)}, 0.4 \textbf{(b)}, 0.6 \textbf{(c)}, 0.7 \textbf{(d)}, 12 \textbf{(e)}, and 80 \textbf{(f)}.
}   
\label{fig_face_equilibria}
\end{figure*}

We assume a large population of agents
that play the game specified by the payoff matrix $A =[a_{ij}]$
and update their strategies by payoff-led social learning.
For instance, an agent can occasionally compare its payoff with that of another agent randomly selected in the population,
 and imitate the strategy of that player if it has a higher payoff.
 Assuming that the probability for the imitation is proportional to the payoff difference,
 the evolution of the frequencies of the strategies in a well-mixed  population is given by the replicator equations
\begin{equation}
\dot{x}_i = x_i \left(\pi_i -\bar{\pi}\right)
= x_i \left(\sum_{j=1}^4 x_j a_{ij} -\sum_{l,j=1}^4 x_l x_j a_{lj}\right),
\label{eq_replicator}
\end{equation}
where the dot denotes 
the
time derivative,
$x_i$
the frequency of the $i$-th strategy,
 with $i\in\{1,2,3,4\}$,
 $\mathbf{x} =\left(x_1, x_2, x_3, x_4\right) =\left(x_{\text{\tiny IT}}, x_{\text{\tiny IU}}, x_{\text{\tiny NT}}, x_{\text{\tiny NU}}\right)$,
$\pi_i$ 
the expected payoff for the $i$-th strategy,
and $\bar{\pi}$ 
the population-mean payoff.
The state space is represented by
the 3-simplex 
$\{(x_1,x_2,x_3,x_4): x_1,x_2,x_3,x_4 \ge 0, x_1 +x_2 +x_3 +x_4 =1\}$.
The replicator dynamics of Eq.\,\eqref{eq_replicator} in a well-mixed population leads to a mixture of $NT$ and $NU$
(equivalent to that of Fig.\,\ref{fig_face_equilibria}a).

\subsubsection{Evolutionary Game Dynamics on Graphs}

For analytical tractability,
we
assume that the network structure is specified by a random regular graph with node degree $k \ge3$,
where the agents occupy the nodes of the graph.
The game interaction and strategy imitation take place 
only between neighbouring agents.
Using the pair approximation 
method originally formulated for an infinitely large Caily tree
that is well approximated by a large (random) regular graph
 \cite{Matsuda:1992ta},
it is shown that the replicator equations on a regular graph are formally equivalent to those in a well-mixed population
with a transformed payoff matrix \cite{Ohtsuki:2006aa}.
Specifically,
for the social learning,
the replicator dynamics on a graph of node degree $k$ with the payoff matrix $A =[a_{ij}]$ is equivalent to
that 
on
a well-mixed population with a payoff matrix $C =[c_{ij}] =[a_{ij}+b_{ij}]$,
where
\begin{equation}
b_{ij} =\frac{(k+3)a_{ii}+3a_{ij}-3a_{ji}-(k+3)a_{jj}}{(k+3)(k-2)}.
\end{equation}
Thus, the replicator equations on a graph are given by
\begin{equation}
\dot{x}_i 
= x_i \left(\sum_{j=1}^4 x_j \left(a_{ij}+b_{ij}\right) -\sum_{l,j=1}^4 x_l x_j \left(a_{lj} +b_{lj}\right)\right).
\label{eq_replicator_graph}
\end{equation}
Due to the condition $\sum_{i=1}^4 x_i =1$,
there are only three independent variables. 
Without loss of generality,
we take $x_1$, $x_2$ and $x_3$ as independent variables. 
 
\section{Equilibria and stability}

To analyse the dynamical system of Eq.\eqref{eq_replicator_graph},
we find all equilibria by solving zero states of it, $\dot{x}_1 =\dot{x}_2 =\dot{x}_3 =0$.
The stability of an equilibrium is analysed with the signs of eigenvalues of the Jacobian matrix $M$ at the equilibrium,
where
\begin{equation}
M
=
\left(
\begin{array}{ccc}
\frac{\partial \dot{x}_1}{\partial x_1} &
\frac{\partial \dot{x}_1}{\partial x_2} &
\frac{\partial \dot{x}_1}{\partial x_3} 
\\
\frac{\partial \dot{x}_2}{\partial x_1}        &
\frac{\partial \dot{x}_2}{\partial x_2}  &
\frac{\partial \dot{x}_2}{\partial x_3}  
\\
\frac{\partial \dot{x}_3}{\partial x_1}        &
\frac{\partial \dot{x}_3}{\partial x_2}  &
\frac{\partial \dot{x}_3}{\partial x_3}  
\end{array}
\right).
\label{eq_jacobian}
\end{equation}

\subsection{1-Morphic Equilibria at the Vertices}
 
\subsubsection{$x_{\text{\tiny IU}}=x_{\text{\tiny NT}}=x_{\text{\tiny NU}}=0$}

The equilibrium $IT$ at the vertex $\left(x_1,x_2,x_3,x_4\right)=\left(x_{\text{\tiny IT}}, x_{\text{\tiny IU}}, x_{\text{\tiny NT}}, x_{\text{\tiny NU}}\right) =(1,0,0,0)$ corresponds to a homogeneous 
state
of the population,
where all 
the agents use the same strategy 
$IT$.
The equilibrium $IT$ can be  asymptotically stable:
trajectories starting close enough to the equilibrium
not only remain close enough but also eventually converge to it.
The Jacobian  $M=[m_{ij}]$ at the equilibrium $IT$ is given by
$m_{11} = \frac{-2 (k+3) r-3 p+6}{k^2+k-6}-p-2 r+1,  m_{12}=\frac{p}{k-2}-r, m_{13}=\frac{3 (p+2 r)-k (k+1) (p+r-1)}{k^2+k-6}$,
$m_{22} =-\frac{k (k (p+r-1)+2 p+3 r-1)}{k^2+k-6}$,
$m_{33} =\frac{k r}{2-k}$,
$m_{21}=m_{23}=m_{31}=m_{32}=0$.
$M$
has the three eigenvalues
$\lambda_{1,a} = -\frac{k r}{k-2}, \lambda_{1,b} =\frac{-k (k+1) (p-1)-2 k (k+2) r+3 (p+2 r)}{k^2+k-6}$ and
$\lambda_{1,c} = -\frac{k (k (p+r-1)+2 p+3 r-1)}{k^2+k-6}$.
If and only if all 
the eigenvalues are negative,
the equilibrium is asymptotically stable.
This condition
is satisfied in the following three cases.
Case $r<\frac{4}{11}$:
$\left(k<k_{1,a}\land p>p_{1,a}\right) \lor
\left(k> k_{1,a}\land p>p_{1,b}\right) \implies \lambda_{1,a}, \lambda_{1,b}, \lambda_{1,c} <0$,
where
$k_{1,a} =\frac{1}{2} \sqrt{\frac{5 r^2+2 r+1}{r^2}}+\frac{1-r}{2 r}$,
$p_{1,a} =\frac{-2 k^2 r+k^2-4 k r+k+6 r}{k^2+k-3}$,
$p_{1,b} =\frac{k (1-r) -3 r+1}{k+2}$,
$\land$ and $\lor$ denote logical `AND' and `OR', respectively.
Case $\frac{4}{11} \le r\leq \frac{2}{3}$ :
$p>p_{1,b}$
(Fig.\,\ref{fig_face_equilibria}d).
Case $\frac{2}{3}<r$:
$\left(k<k_{1,b}\right) \lor \left(k>k_{1,b}\land p>p_{1,b}\right)$,
where
$k_{1,b} =\frac{3r -1}{1-r}$.
As $k\rightarrow\infty$ and, consequently,
$p_{1,b} \rightarrow p_{1,c}=1-r$,
we recover a well-mixed population,
where $IT$ is asymptotically stable for $p > p_{1,c}$.
Since $p_{1,b} <p_{1,c}$,
in this context 
punishment promote the prosocial strategy $IT$ in structured populations more efficiently
than it does
 in a well-mixed population.
If any of the eigenvalues is positive, 
$IT$ is unstable.

\subsubsection{$x_{\text{\tiny IT}}=x_{\text{\tiny NT}}=x_{\text{\tiny NU}}=0$}
The equilibrium $IU =(0,1,0,0)$ is unstable
since one of the eigenvalues is positive,
$\lambda_{2,a} =\frac{(k-1) p}{k-2}+r >0$ (Fig.\,\ref{fig_face_equilibria}a).

\subsubsection{$x_{\text{\tiny IT}}=x_{\text{\tiny IU}}=x_{\text{\tiny NU}}=0$}
The equilibrium $NT =(0,0,1,0)$ is unstable
since one of the eigenvalues is positive,
$\lambda_{3,a} =\frac{k r}{k-2}>0$ (Fig.\,\ref{fig_face_equilibria}a).

\subsubsection{$x_{\text{\tiny IT}}=x_{\text{\tiny IU}}=x_{\text{\tiny NT}}=0$}
The equilibrium $NU =(0,0,0,1)$ 
can be stable. 
It has the eigenvalues
$	\lambda_{4,a} =0, \lambda_{4,b} =-\frac{k (k+p+1)}{k^2+k-6} <0$
and $\lambda_{4,c} =\frac{3 (p+2 r)-k (k-2 r+1)}{k^2+k-6}$.
Since one of the eigenvalues is 0,
$NU$ is not asymptotically stable.
However, it can be (Lyapunov) stable 
if none of the eigenvalues is positive:
trajectories starting close enough to the equilibrium 
remain close enough to it.
We have $p\leq p_4 \implies \lambda_{4,c} \le0 \Longleftrightarrow$ a stable equilibrium,
where $p_4 =\frac{1}{3} \left(k^2-2 k r+k-6 r\right)$ (Fig.\,\ref{fig_face_equilibria}a).
We have $p> p_4 \implies \lambda_{4,c} >0 \Longleftrightarrow$ a unstable equilibrium (Fig.\,\ref{fig_face_equilibria}f).

\begin{figure*}[t]  
\begin{center}  
\includegraphics[width=0.87\textwidth]{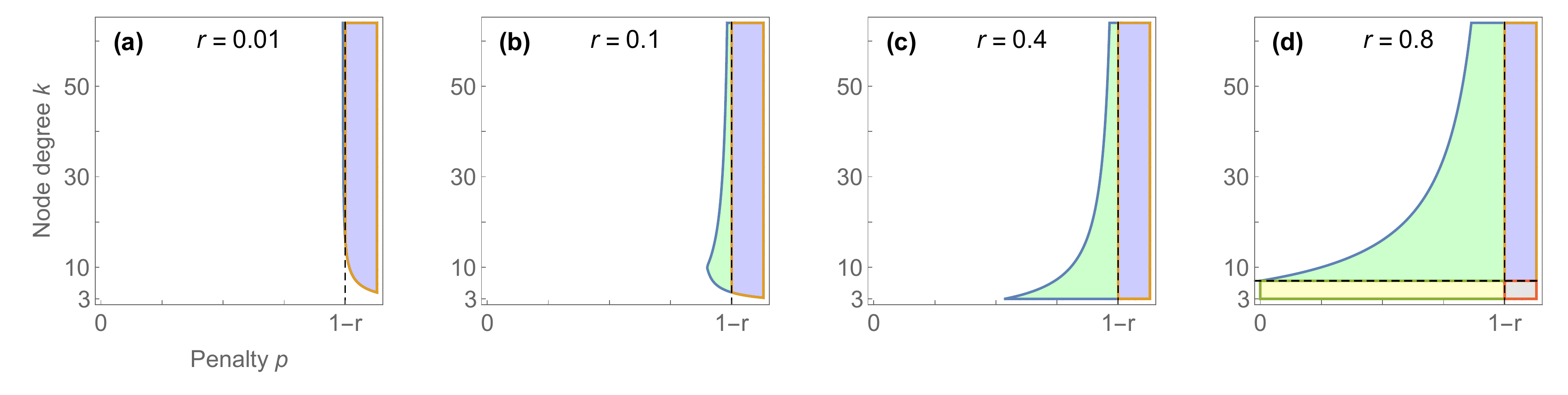}              
\end{center}  
\vspace{-0.3cm}
\caption{
Parameter ranges for the evolution of full trust and trustworthiness.
The shaded regions are the ranges of $(p,k)$ where the equilibrium $IT$ is globally asymptotically stable.
The dashed vertical line indicates $p =p_{1,c} =1-r$:
$p>1-r$ would 
make
$IT$ 
globally asymptotically stable if there is no networked structure (i.e.,\,$k\rightarrow 
\infty$ or a well-mixed population).
The dashed horizontal line indicates $k=k_{1,b}=(3r-1)/(1-r)$:
for $r>2/3$,  
$k< k_{1,b}$  
makes $IT$ 
globally asymptotically stable even with no incentive in a structured population.
Blue: incentive alone ($p>1-r$) would make $IT$ globally asymptotically stable in a well-mixed population and the interplay with networked structure does not make it unstable.
Yellow: networked structure alone ($k <k_{1,b}$) would make $IT$ stable and the interplay with incentives does not make it unstable.
The region in grey is the intersection of those in blue and yellow.	
Green: synergy between incentives and networked structure makes
$IT$ 
globally asymptotically stable
in spite of $p<1-r$  and $k >k_{1,b}$,
each of which would make $IT$ unstable if acting alone.
\textbf{(a)}  
For low $r$,
there is little synergy between the incentive and the networked structure.
Instead, interference between them can occur for low $k$,
where $IT$ is unstable even 
for
large penalty $p>1-r$. 
\textbf{(b)} to \textbf{(d)} As $r$ increases,
synergy emerges and 
its range
expands,
whereas interference recedes.
\textbf{(d)} As $r$ crosses 2/3,
not only the range for synergy further expands
but also networked structure can fully induce network reciprocity for $k <k_{1,b}$, 
making
$IT$
globally asymptotically stable even at $p=0$. 
}   
\label{fig_stable_IT} 
\end{figure*} 

\subsection{2-Morphic Equilibria on the Edges}		

\subsubsection{$x_{\text{\tiny IU}}=x_{\text{\tiny NT}}=0$} 
The equilibria on the $IT$--$NU$ edge
is unstable. 
It can be found by solving $\pi_{\text{\tiny IT}}(x_{\text{\tiny IT,14}},0,0, 1 -x_{\text{\tiny IT,14}}) =\pi_{\text{\tiny NU}}(x_{\text{\tiny IT,14}},0,0, 1 -x_{\text{\tiny IT,14}})$,
which yields
$\mathbf{x}_{\text{\tiny IT--NU}}^* =\mathbf{x}_{1 4}^* =\left(x_{\text{\tiny IT,14}}^*, 0, 0, 1-x_{\text{\tiny IT,14}}^*\right)$,
where $x_{\text{\tiny IT,14}}^* =\frac{k^2-2 (k+3) r+k-3 p}{\left(k^2+k-6\right) (p+2 r)}$.
The condition for the existence of an equilibrium (i.e.,\,$0< x_{\text{\tiny IT,14}}^*<1$) is satisfied in the following cases.
Case $r\leq \frac{1}{8} \left(2+\sqrt{3}\right):$
$p_{1,a}<p<p_4$.
Case $\frac{1}{8} \left(2+\sqrt{3}\right)<r<\frac{1}{2}:$
$k\leq k_{14,a}\land p_{1,a}<p<p_4$,
$ k_{14,a}<k<k_{14,b}\land p<p_4$, 
$ k=k_{14,b} \land 0<p<p_4$,
$  k>k_{14,b}\land p_{1,a}<p<p_4$,
where 
$k_{14,a} =\frac{4 r -1 -\sqrt{64 r^2-32 r+1}}{2 (1-2 r)}$,
$k_{14,b} =\frac{4 r -1 +\sqrt{64 r^2-32 r+1}}{2 (1-2 r)}$.
Case $r=\frac{1}{2}$:
$k=3\land 0<p<2$,
$k\geq 4\land p<\frac{1}{3} \left(k^2-3\right)$.
Case $\frac{1}{2}<r$:
$p<p_4$ (Fig.\,\ref{fig_face_equilibria}a).
The equilibrium is unstable
since one of the eigenvalues is positive:
$\lambda_{14,a} =x_{\text{\tiny IT,14}}^*\left(1 -x_{\text{\tiny IT,14}}^*\right) (p+2r) >0$.

\subsubsection{$x_{\text{\tiny IT}}=x_{\text{\tiny NU}}=0$} 
The equilibrium $\mathbf{x}^*_{\text{\tiny IU--NT}} =\mathbf{x}^*_{23}=\left(0, x_{\text{\tiny IU,23}}^*,1-x_{\text{\tiny IU,23}}^*, 0\right)$ on the $IU$--$NT$ edge 
is unstable,
where $x_{\text{\tiny IU,23}}^* = \frac{1}{2}-\frac{k p}{2 (k-2) (p+2 r)}$.
The condition for the existence of 
an equilibrium (i.e.,\,$0<  x_{\text{\tiny IU,23}}^* <1$) is
$p<  (k -2) r$ (Fig.\,\ref{fig_face_equilibria}a).
Since one of the eigenvalues is negative, $\lambda_{23,a}  = -x_{\text{\tiny IU,23}}^*(1-x_{\text{\tiny IU,23}}^*)(p +2 r) <0$,
the signs of the remaining two eigenvalues determine the stability of the equilibrium.
The sum of the remaining eigenvalues is $\lambda_{23,b} +\lambda_{23,c} =\text{Tr} -\lambda_{23,a}=\frac{(p + 2 r)}{(k-2)} >0$,
where Tr is the trace of the Jacobian matrix.
The equilibrium is unstable
since at least one of the two eigenvalues is positive.

\subsubsection{$x_{\text{\tiny IT}}=x_{\text{\tiny IU}}=0$}

 The $NT$--$NU$ edge is a line of equilibria,
 a part of which can be stable.
$\pi_{\text{\tiny NT}}(0,0,1-x_{\text{\tiny NU}},x_{\text{\tiny NU}}) =\pi_{\text{\tiny NU}}(0,0,1-x_{\text{\tiny NU}},x_{\text{\tiny NU}}) =0$ holds for all $0<x_{\text{\tiny NU}}<1$.
The eigenvalues at an equilibrium $(0,0,1-x_{\text{\tiny NU}},x_{\text{\tiny NU}})$ are
$\lambda_{34,a} =0, \lambda_{34,b}=-\frac{\left(k^2+k-6\right) r (x_{\text{\tiny NU}}-1)+p (k-3 x_{\text{\tiny NU}}+3)+k (k+1) x_{\text{\tiny NU}}}{k^2+k-6}$ and 
$\lambda_{34,c}=\frac{-k (k+1) (r+1) x_{\text{\tiny NU}}+k (k+3) r+3 x_{\text{\tiny NU}} (p+2 r)}{k^2+k-6}$.
Note that $\lambda_{34,b} <\lambda_{34,c}$.
Although the equilibrium cannot be asymptotically stable due to $\lambda_{34,a} =0$,
it is stable 
if and only if $\lambda_{34,b} <0$ and $\lambda_{34,c}\le0$,
which can be satisfied on a part of the line of equilibria as follows.
For $p<p_4$,
$(0, 0, 1 - x_{\text{\tiny NU}}, x_{\text{\tiny NU}})$ is stable with $x_{\text{\tiny NU}} \ge x_{\text{\tiny NU,34}}^*$
and unstable with $x_{\text{\tiny NU}} <x_{\text{\tiny NU,34}}^*$,
where $x_{\text{\tiny NU,34}}^* =\frac{k^2 r+3 k r}{k^2 r+k^2+k r+k-3 p-6 r}$.
For $p>p_4$,
the whole line of equilibria $(0, 0, 1 - x_{\text{\tiny NU}}, x_{\text{\tiny NU}})$ is unstable.

\subsubsection{$x_{\text{\tiny IU}}=x_{\text{\tiny NU}}=0$ } 

There is no equilibrium on the $IT$--$NT$ edge 
since $\pi_{\text{\tiny IT}}\left(x_{\text{\tiny IT}},0,1-x_{\text{\tiny IT}},0\right)  -\pi_{\text{\tiny NT}}(x_{\text{\tiny IT}},0,1-x_{\text{\tiny IT}},0) =\frac{k r}{k-2} >0$,
whereas $\pi_{\text{\tiny IT}}  =\pi_{\text{\tiny NT}}$ should hold at an equilibrium.

\subsubsection{$x_{\text{\tiny IT}}=x_{\text{\tiny NT}}=0$} 

There is no equilibrium on the $IU$--$NU$ edge since
$\pi_{\text{\tiny IU}}(0,x_{\text{\tiny IU}},0,1-x_{\text{\tiny IU}})  -\pi_{\text{\tiny NU}}(0,x_{\text{\tiny IU}},0,1-x_{\text{\tiny IU}})  
    =-\frac{x_{\text{\tiny IU}}\left[k (k+1)  (p+r-1) -3 (p+2 r)\right]+k (k+p+1)}{k^2+k-6} <0
$,
where $0<x_{\text{\tiny IU}}<1$.

\subsubsection{$x_{\text{\tiny NT}}=x_{\text{\tiny NU}}=0$} 
The $IT$--$IU$ edge is a line of equilibria,
which is degenerate.
The condition for the equilibria
$\pi_{\text{\tiny IT}}(x_{\text{\tiny IT}},1-x_{\text{\tiny IT}},0,0) -\pi_{\text{\tiny IU}}(x_{\text{\tiny IT}},1-x_{\text{\tiny IT}},0,0) =\frac{k \left[p(k+2)+(k+3) r-k-1\right]}{k^2+k-6} =0$
is satisfied 
for the whole edge
at 
$p=\frac{-(k+3) r+k+1}{k+2} \land r \le \frac{2}{3}$.
Holding only at a particular value of $p$ for given $r \le \frac{2}{3}$ and $k$,
however,
the line of equilibria is degenerate or structurally unstable,
because
an arbitrarily small perturbation in $p$ leads the line of equilibria to disappear.

\subsection{3-Morphic Equilibria on the Faces}

\subsubsection{$x_{\text{\tiny NU}}=0$}
The equilibrium $\mathbf{x}_{\text{\tiny IT--IU--NT}}^* =\mathbf{x}_{1 2 3}^* =
\left( 1-x_{\text{\tiny IU,123}}^* -x_{\text{\tiny NT,123}}^*,
  x_{\text{\tiny IU,123}}^*,  x_{\text{\tiny NT,123}}^*,    0 
\right)
$
on the $IT$--$IU$--$NT$ face 
is unstable,
where
$  x_{\text{\tiny IU,123}}^*=\frac{k (k+3) r}{k (k+1) (r+1)-3 (p+2 r)}$
and $  x_{\text{\tiny NT,123}}^*=\frac{k (k(p+r-1)+2 p+3 r-1)}{k (k+1) (p+r-1)-3 (p+2 r)}$.	
The equilibrium
is found by solving $\pi_{\text{\tiny IT}}(1-x_{\text{\tiny IU}}-x_{\text{\tiny NT}},x_{\text{\tiny IU}},x_{\text{\tiny NT}},0) =\pi_{\text{\tiny IU}}(1-x_{\text{\tiny IU}}-x_{\text{\tiny NT}},x_{\text{\tiny IU}},x_{\text{\tiny NT}},0) =\pi_{\text{\tiny NT}}(1-x_{\text{\tiny IU}}-x_{\text{\tiny NT}},x_{\text{\tiny IU}},x_{\text{\tiny NT}},0)$.
The conditions for existence of the equilibrium 
(i.e.,\,$0<x_{\text{\tiny IU,123}}^*,  x_{\text{\tiny NT,123}}^*,1 -x_{\text{\tiny IU,123}}^* -x_{\text{\tiny NT,123}}^* <1$)
are satisfied in the following cases.
Case $0<r <\frac{2}{5}$: 
$p_{123}<p<p_{1,b}$,
where   
$p_{123} =\frac{1}{6} (k^3 r+2 k^2 r+k^2-2 k r+k-12 r)
-\frac{1}{6} (k^6 r^2+4 k^5 r^2-20 k^3 r^2-8 k^2 r^2+2 k^5 r+6 k^4 r-12 k^3 r-16 k^2 r+k^4+2 k^3+k^2)^{1/2}$. 
Case    $    \frac{2}{5} < r$:
$k_{1,b}<k<k_{123}\land  p<p_{1,b}$,
$k>k_{123}\land p_{123}<p<p_{1,b}$,
where $k_{123} =\frac{1}{2} \sqrt{\frac{r^2+14 r+1}{(r-1)^2}}+\frac{1-5 r}{2 (r-1)}$.
The equilibrium is unstable
since one of the eigenvalues is positive,
$\lambda_{123,a} =\frac{p+2 r}{k-2} >0$ (Fig.\,\ref{fig_face_equilibria}b).

\subsubsection{$x_{\text{\tiny NT}}=0$}

The equilibrium 
$\mathbf{x}_{\text{\tiny IT--IU--NU}}^* =\mathbf{x}_{\text{124}}^* =
(x_{\text{\tiny IT,124}}^*,
1 -x_{\text{\tiny IT,124}}^* -x_{\text{\tiny NU,124}}^*,
0,
x_{\text{\tiny NU,124}}^*
)
$
on the $IT$--$IU$--$NU$ face 
is unstable,
where
$x_{\text{\tiny IT,124}}^*=\frac{k (k+p+1)}{k (k+1) (r+1)-3 (p+2 r)}$ 
and $x_{\text{\tiny NU,124}}^*=\frac{k (k (p+r-1)+2 p+3 r-1)}{k (k+1) (p+r-1)-3 (p+2 r)} $.
The conditions for existence of the equilibrium 
(i.e.,\,$0<x_{\text{\tiny IT,124}}^*, x_{\text{\tiny NU,124}}^*, 1-x_{\text{\tiny IT,124}}^* -x_{\text{\tiny NU,124}}^*<1$) are satisfied in the following cases.
Case $r\leq \frac{4}{11}$: 
$k>\frac{1}{2} \sqrt{\frac{5 r^2+2 r+1}{r^2}}+\frac{1-r}{2 r}\land p_{124}<p<p_{1,b} 
$,
where $p_{124} =\frac{1}{2}\frac{36 r-k \left((k (2k+5)-15) r+k (k+1)^2\right)}{k \left(k^2+k-6\right)-9}
    +\frac{1}{2}\frac{\sqrt{k^2 (k+3)^2 \left(\left(k^2-4\right) (k+1)^2+(17-4 (k-1) k) r^2-2 (k-2) (k+1) r\right)}}{k \left(k^2+k-6\right)-9}$.
Case $\frac{4}{11}<r\leq \frac{1}{2} \left(\sqrt{17}-3\right)$:  
$p_{124}<p<p_{1,b}$.   
Case $\frac{1}{2} \left(\sqrt{17}-3\right)<r<\frac{2}{3}$:  
$\left(k<k_{124} \land  p<p_{1,b}\right) 
\lor \left(k>k_{124}\land p_{124}<p<p_{1,b}\right)$.
Case $\frac{2}{3}\leq r$: 
$\left(k_{1,b}<k<k_{124} \land  p<p_{1,b}\right) 
\lor (k>k_{124}\land p_{124}<p<p_{1,b})$,
where
$k_{124}$ is the 2nd root of $0=k^4 (r-1) +k^3 \left(2 r^2+4 r-2\right)+k^2 \left(8 r^2+3 r-1\right)  -6 k r^2-36 r^2
$.

Although the first eigenvalue $\lambda_{124,a} =-\frac{p+2 r}{k-2} <0$  is negative,
the sum of remaining two  eigenvalues is
$\lambda_{124,b} + \lambda_{124,c} 
=\text{Tr} -\lambda_{124,a}
 =\frac{k^2 (k+p+1) (p+2 r)\left[(k+2) p+k (r-1)+3 r-1\right]}{\left[-3 p +k (k+1) (r+1) -6 r\right] 
		 \left[\left(k (1 + k)-3\right) p  + k (1 + k) (r-1)- 6 r \right]} > 0
$		 
and, thus, one of the (real parts of) two eigenvalues is positive.
Hence, the equilibrium is unstable
(Fig.\,\ref{fig_face_equilibria}c).

\subsubsection{$x_{\text{\tiny IU}}=0$}
The equilibria on the $IT$--$IU$--$NT$ face are degenerate.
A line of equilibria exists only at a particular value of $p=\frac{6 r-k (k+1) (r-1)}{k^2+k-3}$, given $r$ and $k$.

\subsubsection{$x_{\text{\tiny IT}}=0$}
The equilibria on the $IU$--$NT$--$NU$ face are degenerate
for the same reason as above.

\subsection{No 4-Morphic or Interior Equilibrium}
There is no interior equilibrium.
For an interior point $(x_{\text{\tiny IT}}, x_{\text{\tiny IU}},x_{\text{\tiny NT}},x_{\text{\tiny NU}})$, i.e.,\,$0<x_{\text{\tiny IT}}, x_{\text{\tiny IU}},x_{\text{\tiny NT}},x_{\text{\tiny NU}}<1$,
we have $\pi_{\text{\tiny NT}}(x_{\text{\tiny IT}}, x_{\text{\tiny IU}},x_{\text{\tiny NT}},x_{\text{\tiny NU}})  \ne \pi_{\text{\tiny NU}}(x_{\text{\tiny IT}}, x_{\text{\tiny IU}},x_{\text{\tiny NT}},x_{\text{\tiny NU}})$
since $\pi_{\text{\tiny NT}} -\pi_{\text{\tiny NU}} 
=-\frac{(x_{\text{\tiny IT}}+x_{\text{\tiny IU}}) (k (k+1) (1-r-p) +3 (p+2 r))}{(k-2)(k+3)} <0$
for $p \le 1-r$ 
and
$\pi_{\text{\tiny NT}} -\pi_{\text{\tiny NU}} 
= [p ((k+3) w+(k+3) z)+(k+3) r (2 w+2 z) +k(x_{\text{\tiny IT}}+x_{\text{\tiny IU}})((p+r-1)k+2p-3r-1)] / [(k-2)(k+3)] >0
$
for $p>1-r$.
Hence, no interior point
satisfies the condition 
$\pi_{\text{\tiny IT}} =\pi_{\text{\tiny IU}} =\pi_{\text{\tiny NT}} =\pi_{\text{\tiny NU}}$
for an equilibrium.

In general,
replicator dynamics of a normal-form or matrix-form game with four strategies 
can have steady states (e.g.,\,a limit cycle or a chaotic attractor)
other than an isolated equilibrium point in the interior state space.
Since the dynamical system of Eq.\,\eqref{eq_replicator_graph} contains no interior equilibrium,
however, there exist no steady states in the interior state space, according to Theorem 7.6.1 of the reference \cite{hofbauer1998evolutionary}.

\section{Interference and Synergy}

\begin{figure}[t]  
\begin{center}     
\includegraphics[width=0.46\textwidth]{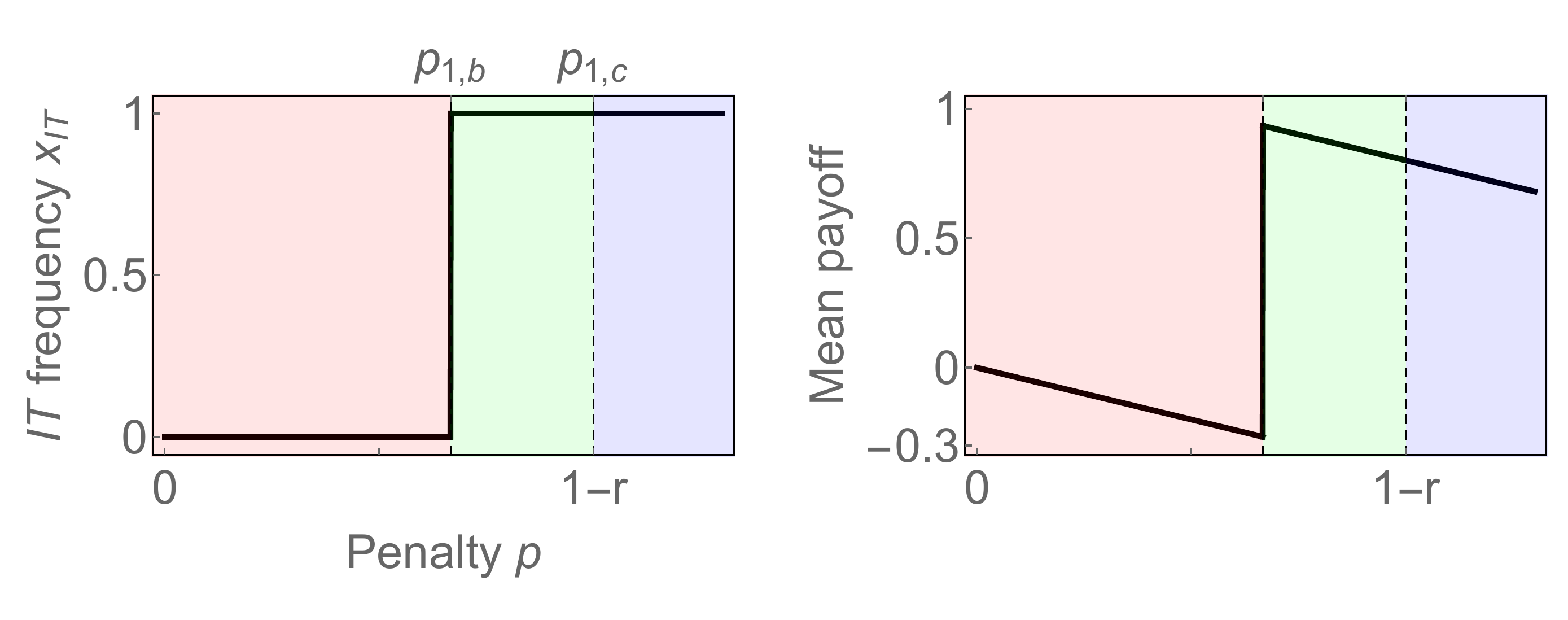}     
\end{center}  
\vspace{-0.3cm}
\caption{
The frequency $x_{\text{\tiny IT}}$ of the 
prosocial 
strategy $IT$ and population-mean payoff
at the evolutionary stable state
in terms of penalty size $p$ 
	\fxwarning{
	(normalised by $1-r$)  
	}
in a structured population.
While the evolution of $IT$ is maintained for $p\ge p_{1,b}$,
the mean payoff is optimal at $p= p_{1,b}$.
Parameters: $r=0.6, k=10, f=p$.
}   
\label{fig_payoff}
\end{figure}

Our analysis shows that punishment and 
random regular graphs
interact in a non-trivial way.
For low $r$, 
interference can occur at low node degrees $k$:
this prevents the evolution of the prosocial strategy
$IT$ even at a high level of penalty $p>p_{1,c}$ 
that would be sufficient if 
the evolution
were 
on
a well-mixed population (Fig.\,\ref{fig_stable_IT}a).
As $r$ increases,
however, 
interference recedes
whereas synergy emerges and 
the range of it
expands:
a combination of even low penalty and weakly networked structure (i.e.,\,large degrees)
can lead to the evolution of $IT$,
each of which would fail if acting alone
(Fig.\,\ref{fig_stable_IT}b to \ref{fig_stable_IT}d).
The synergy not only lowers the level of penalty $p=p_{1,b}$ required for the evolution of $IT$
but also yields a higher payoff 
than penalty alone $p=p_{1,c}$ does in a well-mixed population
(Fig.\,\ref{fig_payoff}).

We have shown that simple networks 
are sufficient  
to yield a substantial interplay with incentives
for promoting pro-social behaviours
in large multi-agent systems.
For future work,
impacts of 
complex networks,
stochastic game dynamics, 
interplays between other mechanisms in TG 
and other 
games involving pro-sociality
will also be well worth studying.

\section*{Acknowledgment}
The authors would like to thank Naoki Masuda 
for helpful comments.

\bibliographystyle{IEEEtran} 

\begin{thebibliography}{10}
\providecommand{\url}[1]{#1}
\csname url@samestyle\endcsname
\providecommand{\newblock}{\relax}
\providecommand{\bibinfo}[2]{#2}
\providecommand{\BIBentrySTDinterwordspacing}{\spaceskip=0pt\relax}
\providecommand{\BIBentryALTinterwordstretchfactor}{4}
\providecommand{\BIBentryALTinterwordspacing}{\spaceskip=\fontdimen2\font plus
\BIBentryALTinterwordstretchfactor\fontdimen3\font minus
  \fontdimen4\font\relax}
\providecommand{\BIBforeignlanguage}[2]{{%
\expandafter\ifx\csname l@#1\endcsname\relax
\typeout{** WARNING: IEEEtran.bst: No hyphenation pattern has been}%
\typeout{** loaded for the language `#1'. Using the pattern for}%
\typeout{** the default language instead.}%
\else
\language=\csname l@#1\endcsname
\fi
#2}}
\providecommand{\BIBdecl}{\relax}
\BIBdecl

\bibitem{Zhang:2020aa}
J.~Zhang and M.~Cao, ``Strategy competition dynamics of multi-agent systems in
  the framework of evolutionary game theory,'' \emph{IEEE Transactions on
  Circuits and Systems II: Express Briefs}, vol.~67, no.~1, pp. 152--156, 2020.

\bibitem{Abbass:2016aa}
H.~Abbass, G.~Greenwood, and E.~Petraki, ``The $n$-player trust game and its
  replicator dynamics,'' \emph{IEEE Transactions on Evolutionary Computation},
  vol.~20, no.~3, pp. 470--474, 2016.

\bibitem{Lim:2020aa}
\BIBentryALTinterwordspacing
I.~S. Lim, ``Stochastic evolutionary dynamics of trust games with asymmetric
  parameters,'' \emph{Physical Review E}, vol. 102, no.~6, pp. 062\,419--, 12
  2020.
\BIBentrySTDinterwordspacing

\bibitem{nowak2006five}
\BIBentryALTinterwordspacing
M.~A. Nowak, ``Five rules for the evolution of cooperation,'' \emph{Science},
  vol. 314, no. 5805, pp. 1560--1563, 12 2006.
\BIBentrySTDinterwordspacing

\bibitem{Sasaki:2012rz}
\BIBentryALTinterwordspacing
T.~Sasaki, {\AA}.~Br{\"a}nnstr{\"o}m, U.~Dieckmann, and K.~Sigmund, ``The
  take-it-or-leave-it option allows small penalties to overcome social
  dilemmas,'' \emph{Proceedings of the National Academy of Sciences}, vol. 109,
  no.~4, p. 1165, 01 2012.
\BIBentrySTDinterwordspacing

\bibitem{Capraro:tw}
\BIBentryALTinterwordspacing
V.~Capraro and M.~Perc, ``Mathematical foundations of moral preferences,''
  \emph{Journal of The Royal Society Interface}, vol.~18, no. 175, p. 20200880,
  2021.
\BIBentrySTDinterwordspacing

\bibitem{Perc:2017aa}
\BIBentryALTinterwordspacing
M.~Perc, J.~J. Jordan, D.~G. Rand, Z.~Wang, S.~Boccaletti, and A.~Szolnoki,
  ``Statistical physics of human cooperation,'' \emph{Physics Reports}, vol.
  687, pp. 1--51, 2017.
\BIBentrySTDinterwordspacing

\bibitem{Sigmund:2010aa}
\BIBentryALTinterwordspacing
K.~Sigmund, H.~De~Silva, A.~Traulsen, and C.~Hauert, ``Social learning promotes
  institutions for governing the commons,'' \emph{Nature}, vol. 466, no. 7308,
  pp. 861--863, 2010.
\BIBentrySTDinterwordspacing

\bibitem{Zhang:2019aa}
C.~Zhang, Q.~Li, Y.~Zhu, and J.~Zhang, ``Dynamics of task allocation based on
  game theory in multi-agent systems,'' \emph{IEEE Transactions on Circuits and
  Systems II: Express Briefs}, vol.~66, no.~6, pp. 1068--1072, 2019.

\bibitem{taylor1978evolutionary}
\BIBentryALTinterwordspacing
P.~D. Taylor and L.~B. Jonker, ``Evolutionary stable strategies and game
  dynamics,'' \emph{Mathematical Biosciences}, vol.~40, no. 1--2, pp. 145 --
  156, 1978.
\BIBentrySTDinterwordspacing

\bibitem{Ohtsuki:2006fk}
\BIBentryALTinterwordspacing
H.~Ohtsuki, C.~Hauert, E.~Lieberman, and M.~A. Nowak, ``A simple rule for the
  evolution of cooperation on graphs and social networks,'' \emph{Nature}, vol.
  441, no. 7092, pp. 502--505, 05 2006.
\BIBentrySTDinterwordspacing

\bibitem{Chica:2018wt}
M.~Chica, R.~Chiong, M.~Kirley, and H.~Ishibuchi, ``A networked n-player trust
  game and its evolutionary dynamics,'' \emph{IEEE Transactions on Evolutionary
  Computation}, vol.~22, no.~6, pp. 866--878, 2018.

\bibitem{Ohtsuki:2006aa}
\BIBentryALTinterwordspacing
H.~Ohtsuki and M.~A. Nowak, ``The replicator equation on graphs,''
  \emph{Journal of Theoretical Biology}, vol. 243, no.~1, pp. 86--97, 2006.
\BIBentrySTDinterwordspacing

\bibitem{Pacheco:2006ve}
\BIBentryALTinterwordspacing
J.~M. Pacheco, A.~Traulsen, and M.~A. Nowak, ``Coevolution of strategy and
  structure in complex networks with dynamical linking,'' \emph{Physical Review
  Letters}, vol.~97, no.~25, pp. 258\,103--, 12 2006.
\BIBentrySTDinterwordspacing

\bibitem{Fang:2021ut}
\BIBentryALTinterwordspacing
X.~Fang and X.~Chen, ``Evolutionary dynamics of trust in the n-player trust
  game with individual reward and punishment,'' \emph{The European Physical
  Journal B}, vol.~94, no.~9, p. 176, 2021.
\BIBentrySTDinterwordspacing

\bibitem{Dong:2019aa}
\BIBentryALTinterwordspacing
Y.~Dong, T.~Sasaki, and B.~Zhang, ``The competitive advantage of institutional
  reward,'' \emph{Proceedings of the Royal Society B: Biological Sciences},
  vol. 286, no. 1899, p. 20190001, 2019.
\BIBentrySTDinterwordspacing

\bibitem{Johnson:2011aa}
\BIBentryALTinterwordspacing
N.~D. Johnson and A.~A. Mislin, ``Trust games: A meta-analysis,'' \emph{Journal
  of Economic Psychology}, vol.~32, no.~5, pp. 865--889, 2011.
\BIBentrySTDinterwordspacing

\bibitem{Tarnita:2015aa}
C.~Tarnita, ``Fairness and trust in structured populations,'' \emph{Games},
  vol.~6, no.~3, pp. 214--230, 2015.

\bibitem{Kumar:2020aa}
\BIBentryALTinterwordspacing
A.~Kumar, V.~Capraro, and M.~Perc, ``The evolution of trust and
  trustworthiness,'' \emph{Journal of The Royal Society Interface}, vol.~17,
  no. 169, p. 20200491, 2020.
\BIBentrySTDinterwordspacing

\bibitem{Masuda:2012aa}
N.~Masuda and M.~Nakamura, ``Coevolution of trustful buyers and cooperative
  sellers in the trust game,'' \emph{PloS one}, vol.~7, no.~9, p. e44169, 2012.

\bibitem{Cho:2015aa}
J.-H. Cho, K.~Chan, and S.~Adali, ``A survey on trust modeling,'' \emph{ACM
  Computing Surveys}, vol.~48, no.~2, pp. 28:1--40, 2015.

\bibitem{Jung:2019yq}
T.~Jung, X.~Li, W.~Huang, Z.~Qiao, J.~Qian, L.~Chen, J.~Han, and J.~Hou,
  ``Accounttrade: Accountability against dishonest big data buyers and
  sellers,'' \emph{IEEE Transactions on Information Forensics and Security},
  vol.~14, no.~1, pp. 223--234, 2019.

\bibitem{Niyato:2009dq}
D.~Niyato, E.~Hossain, and Z.~Han, ``Dynamics of multiple-seller and
  multiple-buyer spectrum trading in cognitive radio networks: A game-theoretic
  modeling approach,'' \emph{IEEE Transactions on Mobile Computing}, vol.~8,
  no.~8, pp. 1009--1022, 2009.

\bibitem{Matsuda:1992ta}
\BIBentryALTinterwordspacing
H.~Matsuda, N.~Ogita, A.~Sasaki, and K.~Sat{\=o}, ``Statistical mechanics of
  population: The lattice lotka-volterra model,'' \emph{Progress of Theoretical
  Physics}, vol.~88, no.~6, pp. 1035--1049, 1992.
\BIBentrySTDinterwordspacing

\bibitem{hofbauer1998evolutionary}
J.~Hofbauer and K.~Sigmund, \emph{Evolutionary Games and Population
  Dynamics}.\hskip 1em plus 0.5em minus 0.4em\relax Cambridge University Press,
  1998.

\end{thebibliography}


\end{document}